# The Enhanced Performance of $C_2H_2$ Gas Sensor based on Carbon-decorated $SnO_2$ Nanoparticles via CVD method

Chunhong Luan, Bao Xu, Chao Wang

*Clean Energy Materials and Engineering Center, School of Microelectronics and Solid-State Electronics, University of Electric Science and Technology of China, Chengdu 611731, P.R. China*

**Abstract**

As a flammable and explosive industrial gas, $C_2H_2$ gas detection is necessary and significant. Herein, highly sensitive $C_2H_2$ gas sensor was initially fabricated by carbon-decorated $SnO_2$ nanoparticles, which was prepared using CVD method with $C_2H_2$ Gas as raw material. The effects of preparation conditions on sensor response were studied and the response to 1000 ppm $C_2H_2$ at working temperature of 370 °C reached a value of 106.3, which is much higher than that of other materials reported. In addition, the carbon-decorated $SnO_2$ sensor showed very good stability and $C_2H_2$ selectivity against other gases. Raman spectroscopy reveals that the response of sensors is affected by carbon-loading and degree of graphitization, which depend on the CVD temperature. It is speculated that the sensor response is enhanced by electron penetration effect from internal $SnO_2$ nanoparticles to the external carbon layer, due to Schottky Barrier formed between $SnO_2$ nanoparticle and carbon layer.

## 1. Introduction

Acetylene ($C_2H_2$) is a colorless, odorless, toxic, flammable and explosive commonly industrial gas, so rapid, sensitive and accurate detection of $C_2H_2$ has very important significance. Over the past decades, the $C_2H_2$ gas sensors are mostly based on ZnO [1-3], $SnO_2$ [4, 5], carbon nanotubes[6, 7], fiber-optic [8, 9], MIS[10] and others. N. Tamaekong prepared Pt/ZnO thick by flame spray pyrolysis (FSP) technique, and the sensors based on 2 at.% Pt loaded ZnO film showed an optimum $C_2H_2$ response of ~836 at 1% acetylene concentration and 300 °C operating temperature [2]. W. Chen et al studied the influence of $SnO_2$ nanostructure morphology on the sensing performances for $C_2H_2$ detection, and they found the gas response of $SnO_2$ nanowires to 1000 ppm $C_2H_2$ reached saturation about 88 at 250 ºC[4]. Gold nanoparticles supported on multi-walled carbon nanotubes (Au/MWCNTs) were prepared by C. Li et al. for an acetylene sensor, which showed a good sensitivity of 80nA/ppm and a response time of 25 s to 50 ppm acetylene [6]. Fiber-optic acetylene gas sensor based on a microstructured optical fiber Bragg grating with a length of 6.5 cm has the absorption to $C_2H_2$ is ~0.13 dB at the concentration of 3.5%, but the response time is as long as ~12 min[8]. In these studies, metal oxide semiconductor gas sensors have attract great attention, due to its strong heat resistance, corrosion resistance, low cost, simple fabrication process and portability.

[*]Corresponding authors. Tel.: +86 28 61831326; fax: +86 28 61831326. E-mail address: cwang@uestc.edu.cn (C. Wang).

As an n-type semiconductor with a wide band gap ($Eg = 3.6$ eV), $SnO_2$ is an old material extensively used to fabricate various gas sensors [11-16]. In order to improve the metal oxide semiconductor sensor response, many researches are usually focus on the following two aspects: doping with catalyst or altering the morphologies. The catalyst doped into $SnO_2$ gas sensors are almost noble metals, such as Pt, Ag, Au, which make the sensors show very high gas response. The reported results [1, 2, 5] indicate that modifying the surface of oxide semiconductors with noble metal is a very effective approach to prepare highly sensitive sensors. However, in spite of the improvement of gas response, the high cost limits the application of these gas sensors. On the other hand, altering the morphologies and structures can increase the surface area and the active sites of gas materials, so it can improve the gas response value of sensor to limited extent [4, 17, 18]. Moreover, these fabrication methods still have some disadvantages. For example, many factors in the preparation process are difficult to be precisely controlled, resulting in poor reproducibility of the sensing properties. In addition, the synthesis of $SnO_2$ gas sensitive materials with special structures requires either complicated preparation process or sophisticated equipments. For gas sensors, all the factors mentioned above limits its wide range of application.

Recently, carbon-decorated or carbon-coated $SnO_2$ nanoparticles were prepared by various methods, and were used as Li-ion battery anodes [19, 20] and catalysts of oxygen reduction reaction [21-23], which show very high performance. Herein, the chemical vapor deposition (CVD) method was used to prepare carbon-decorated $SnO_2$ nanoparticles as a kind of gas sensitive material, and then gas sensors based on this material was fabricated. The sensors based on this materials exhibit high gas response and selectivity to $C_2H_2$ gas, and the related sensing mechanism was also tried to explain.

**2. Experimental**

*2.1 Treating processes of $SnO_2$ nanoparticles*

The starting materials were $SnO_2$ nanoparticles (with average size of 50–70 nm, purchased from Aladdin). $C_2H_2$ gas and argon gas, with purity of 99.99%, was also purchased from Messer gas products Co. Ltd. At first, 3 g $SnO_2$ nanoparticles were placed in a vacuum tube furnace. Then, at a speed of 5 $^{o}C \cdot min^{-1}$, the temperature was increased to 200, 300, 400, 500 and 600 $^{o}C$, respectively, and kept constant for 30 min. After that, 30 $ml \cdot min^{-1}$ $C_2H_2$ gas mixed with 60 $ml \cdot min^{-1}$ argon gas was injected into the tube furnace, and kept for 5 min. Thus, the carbon-decorated $SnO_2$ nanoparticles was obtained, and then they were used to fabricate gas sensors.

*2.2 Fabrication of gas sensor from carbon-decorated $SnO_2$ nanoparticles*

1 g carbon-decorated $SnO_2$ nanoparticles above was ground in an agate mortar for 15 min, respectively, then 5 ml of de-ionized water was added to the ground $SnO_2$ nanoparticles, and the mixture was ground for another 15 min. The resulting paste was coated on the outer surface of an $Al_2O_3$ tube with a pair of Au electrodes attached with Pt lead wires. After that, the $SnO_2$ nanoparticles on the $Al_2O_3$ tube were sintered at 500 $^{o}C$ in the argon gas with the flow rate of 60 $ml \cdot min^{-1}$ for 2 h. Finally, Sensor S1, S2, S3, S4 and S5 were obtained. For comparison purposes, the raw material $SnO_2$ nanoparticles were also fabricated gas sensor in the same way and named as S0.

*2.3 Characterization of the samples*

Gas-sensing performance of $SnO_2$ sensors was tested on HANWEI HW-30A equipment (Hanwei Electronics Co. Ltd., Henan, China). Prior to the measurement, all the sensors were aged in air at 300 $^{o}C$ for 72 h. SEM images of the sample were taken on a HITACHI S3400+EDX field-emission scanning electron microscope. High resolution transmission electron microscopy (HRTEM, JEOL – 2100, the accelerating voltage of 200 kV) was characterized and analyzed. The sample was firstly treated by ultrasonic dispersion in ethanol, and then was added on the copper grid. The XRD patterns were obtained using XRD-7000 X-ray powder diffract meter. The Raman spectroscopy is measured on Lab Ram HR 800 Raman spectrometer produced by France HORIBA Jobin Yvon Company, with laser wavelength 532 nm and operating power 25 mW.

## 3. Results and Discussion

*3.1 Characterization of sensing material*

The SEM images of carbon-decorated $SnO_2$ nanoparticles prepared at different temperatures are shown in the Fig.1. It can be seen that the nanoparticles have the uniform size of 50-70 nm and good dispersion, and there is so little difference in the micro morphology of all the samples. It suggests that the carbon-decorating processing does not have an apparent impact on the morphology of $SnO_2$ nanoparticles, and the carbon deposited on the $SnO_2$ nanoparticles is so thin that it cannot distinguish from the SEM images.

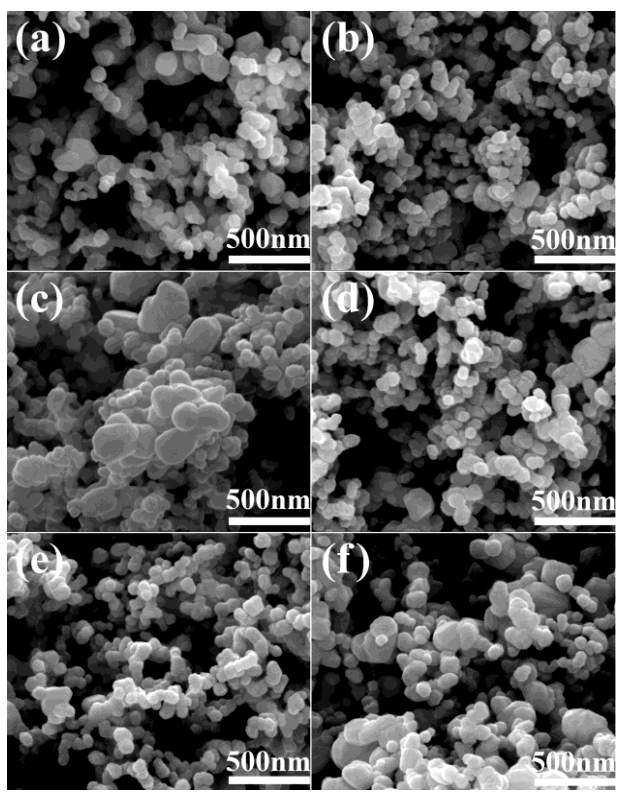

Fig. 1 SEM images of carbon--decorated $SnO_2$ nanoparticles prepared at different

temperatures: (a) S0, (b) S1-200 °C, (c) S2-300 °C, (d) S3-400 °C, (e) S4-500 °C and (f) S5-600 °C.

Further, as a representative, high resolution TEM image of carbon-decorated $SnO_2$ nanoparticles S3 is given in Fig. 3. It can be seen clearly that the $SnO_2$ nanoparticles are coated by carbon layers, of which the crystal lattice is obviously different from that of $SnO_2$ nanoparticle. It illustrates that, the decomposition of $C_2H_2$ gas happens and it turned into carbon layer depositing on the surface of $SnO_2$ nanoparticles in the thermal and anaerobic environment during the CVD process.

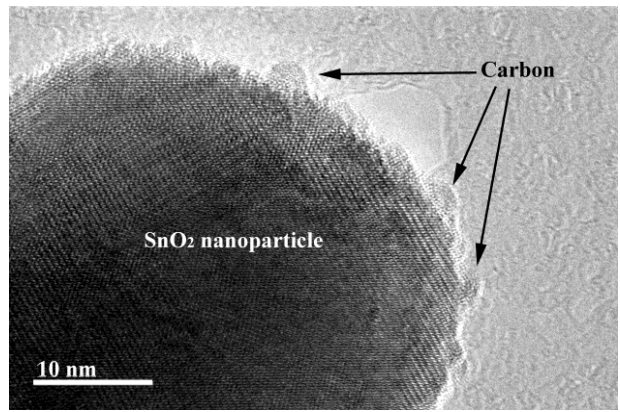

Fig. 2 TEM image of carbon-decorated $SnO_2$ nanoparticles S3

In order to analyze the evolution in crystal structure of the sensing material, the XRD patterns of S1, S2, S3, S4 and S5 are presented in Fig. 3. All the peaks in Fig.3 can be indexed to $SnO_2$ (PDF No.: 41-1445), and there is no obvious peak of carbon observed. It shows the carbon deposited on the $SnO_2$ nanoparticle surface is amorphous. In additional, no apparent change has been observed after comparing these XRD patterns. This phenomenon suggests that the phase structure and average particle size of $SnO_2$ nanoparticles remain unchanged during the treating process.

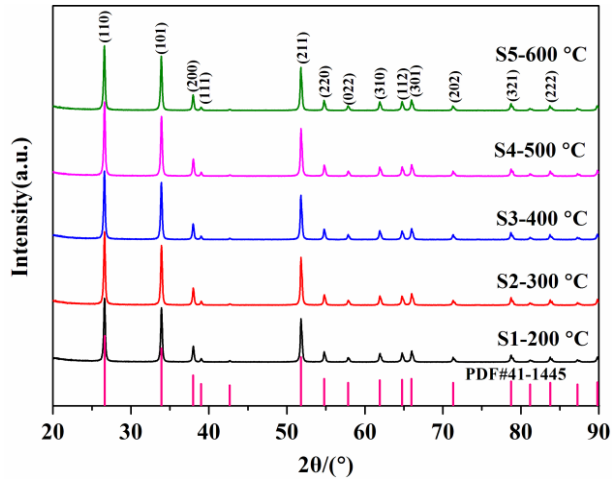

Fig. 3 The XRD patterns of Sample S1, S2, S3, S4 and S5.

*3.2 Gas-sensing performance of carbon-decorated SnO$_2$ nanoparticle Sensors*

In our experiment, the sensor response *S* is defined as *S* = Ra/Rg, where Ra and Rg are the resistances of a sensor in air and in the target gas, respectively. Following this definition, the response of sensor S0, S1, S2, S3, S4 and S5 to 1000 ppm $C_2H_2$ as a function of working temperature ($T_w$) is presented in Fig. 4, which clearly shows that the response of all the sensors increases with $T_w$ increasing at first, up to the maximum value at 370 °C, then decreases with $T_w$ further increasing. The inset of Fig.4 gives the variation of all the sensor response at the optimum working temperature, namely $T_w$ is 370 °C. Comparing to sensor S0, the response of the other sensors increases with the preparation temperature ($T_p$), and reaches its maximum value of 106.3 at 400 °C, then decreases when $T_p$ rose continually.

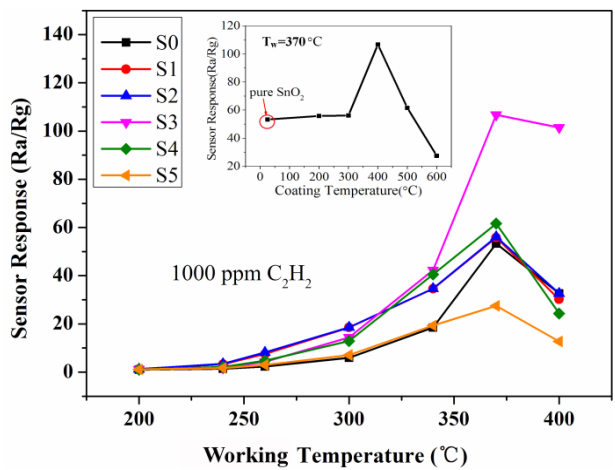

Fig. 4 Relationship between the sensor response and working temperature. The inset is the relationship between the sensor responses and preparation temperature

at working temperature of 370 °C.

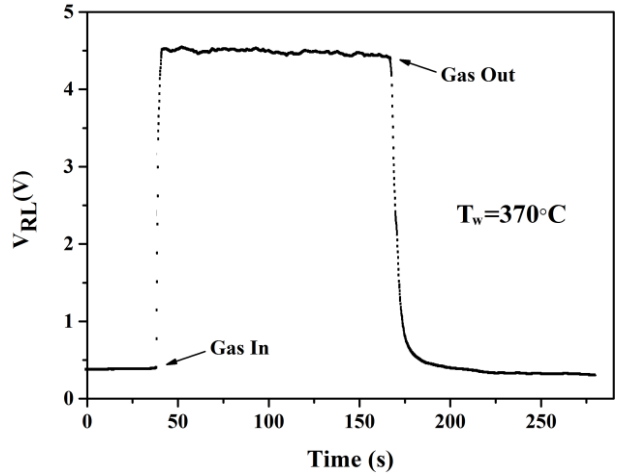

Fig. 5 Curves of response signal voltage of sensor S3 vs. time. The curve is recorded at $T_w$=370 °C in 1000 ppm $C_2H_2$ when $C_2H_2$ was injected into and evacuated out of the chamber.

Fig. 5 shows the variation of response curve at 370 °C upon injecting and evacuating $C_2H_2$. $C_2H_2$ was injected into the chamber at 40 s and was evacuated at 165 s. Generally, the response time is defined as the duration within which the response signal reaches 90% of its maximum value after $C_2H_2$ was injected into the chamber, and the recovery time is the interval in which the signal voltage decreases by 90% of its maximum value when $C_2H_2$ was evacuated. According to this definition, the response time of sensor S3 to 1000 ppm $C_2H_2$ is estimated to be 3 s, while the recovery time is 9 s.

For the gas sensors, selectivity is another key parameter. Fig. 6 shows the response of sensor S3 to various gases at 370 °C. The concentrations of $C_2H_2$, $H_2$ and $CH_4$, are all the same, namely 1000 ppm, and that of $C_2H_5OH$ is 100 ppm. Obviously, sensor S3 exhibits much higher response to $C_2H_2$ than to the other gases, which indicates its quite good selectivity.

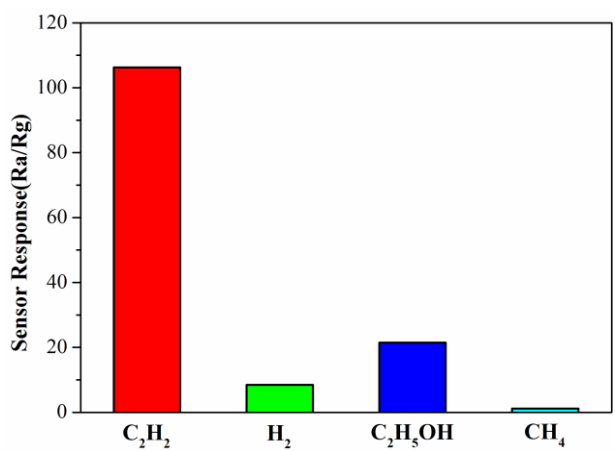

Fig. 6 The selectivity of sensor S3. The concentrations of $C_2H_2$, $H_2$ and $CH_4$, are all 1000 ppm, and that of $C_2H_5OH$ is 100 ppm.

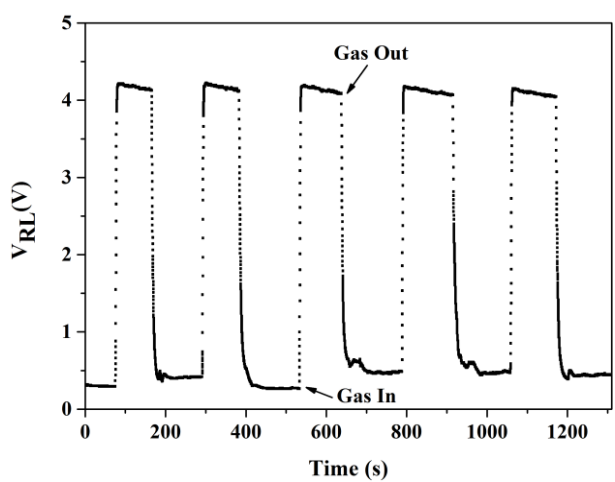

Fig. 7 The stability of sensor S3. $C_2H_2$ gas of 1000 ppm was repeatedly injected into and evacuated from the testing-chamber, and the sensor response remains unchanged within 1200 s.

On the other hand, Fig. 7 presents the variation of response signal of S3 upon repeatedly injecting and evacuating $C_2H_2$ gas within 1200 s. It is clear that sensor response remains unchanged and the response curve can be repeatedly obtained after many operation cycles, revealing that the stability and reproducibility of carbon-decorated $SnO_2$ sensor is rather good and can be used as recyclable gas sensor.

*3.3 Analysis of the sensing mechanism of carbon-decorated $SnO_2$ sensors*

The gas sensor based on carbon-decorated $SnO_2$ shows good gas response to $C_2H_2$ gas. The mechanism is assumed to be attributed to the electron penetration from $SnO_2$ nanoparticle core to the external surface of carbon layer. It is reported that the carbon-decorated metal catalyst and metal oxide will enhance electron penetration through an ultrathin carbon for highly efficient catalysis of the oxygen reduction reaction [22], because the Schottky Barrier may be formed between $SnO_2$ nanoparticle and carbon layer. The effect can induce the electron density of the surface of the carbon-decorated $SnO_2$ nanoparticles is higher than that of pure $SnO_2$ nanoparticles. Higher electron density can absorb more $O_2$, and more absorbed $O_2$ can be activate to $O_2^-$, $O^-$ or $O^{2-}$[17], which can react with reducing gas, for example $C_2H_2$. As a result, the gas response of sensor to $C_2H_2$ will become higher. The carbon-decorated $SnO_2$ nanoparticles, on which $O_2$ can be readily activated by the electrons transferred from the $SnO_2$ nanoparticle to the carbon surface, has recently been demonstrated as a promising strategy to produce robust non-precious gas sensing materials.

However, the phase structure of carbon might affect the process of the electron transfer, and subsequently the oxygen-reduction-reaction activity as the electronic structure of the outermost carbon layer is only modulated by the electron transferred from the $SnO_2$ nanoparticle[24]. Here, the Raman spectra at 1050-1750 cm$^{-1}$ and peak fitting results of $SnO_2$ nanoparticles decorated by carbon under different temperatures are shown in Fig. 8 and Table 1. The spectra are divided into two peaks: D and G, where D peak usually correspond to the defects and disordered carbon impurity on the graphite layer, and the G peak usually represent the graphite structure of carbon decorating. The I(D)/I(G) ratio of the two peak height, is often used to represent the order degree of carbon surface structure, namely graphitization degree. Table 1 shows D peak height, G peak height and I(D)/I(G) ratio. We can find that the I(D)/I(G) ratio increases from 0.56 to 0.88 with the increase of $T_p$. It suggests that the higher $T_p$ can get better graphitization degree of carbon layer, which play the dominant role in the oxygen-reduction-reaction of gas sensitive response.

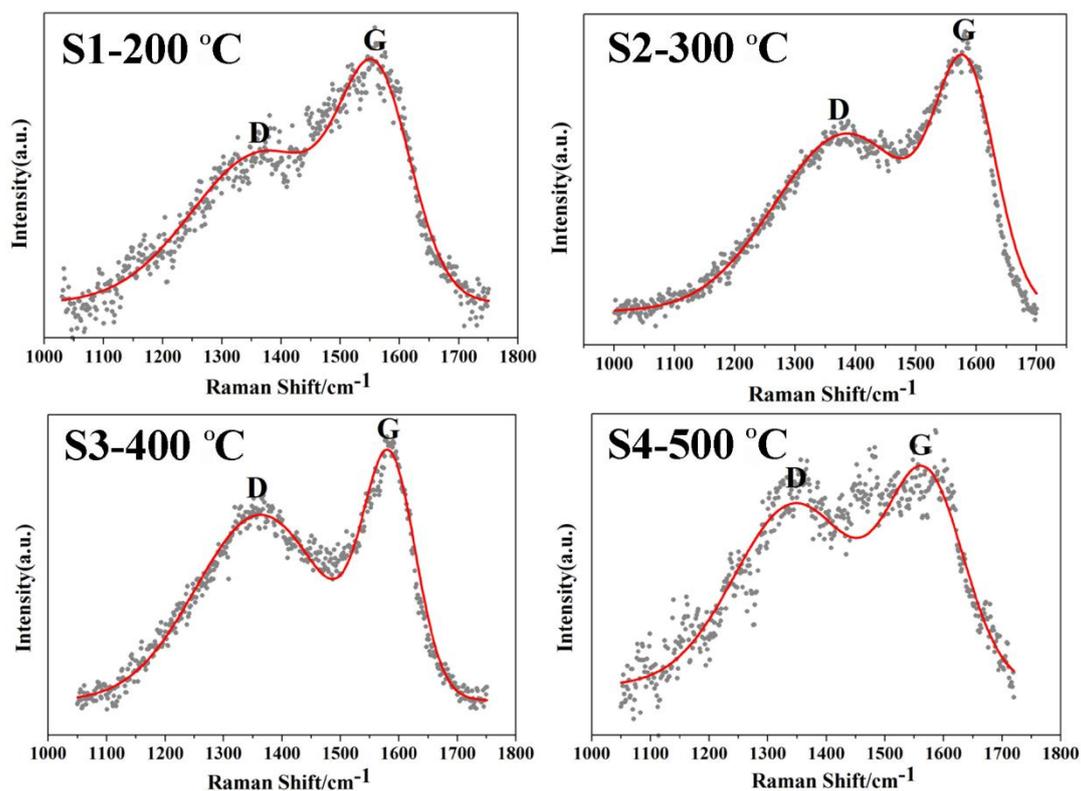

Fig. 8 Raman spectra at 1050-1750 cm$^{-1}$ of Sample S1, S2, S3 and S4

Table 1 I(D)/I(G) ratio of Raman spectra of Sample S1, S2, S3 and S4

| Samples | D peak height | G peak height | I(D)/I(G) |
|---------|---------------|---------------|-----------|
| S1 | 112.74 | 199.57 | 0.56 |
| S2 | 562.89 | 832.72 | 0.68 |
| S3 | 276.36 | 355.71 | 0.78 |
| S4 | 97.429 | 109.97 | 0.88 |

As for the SnO$_2$/carbon composite material, the graphitized carbon with $sp^2$ hybridization has a positive effect in the oxygen reduction reaction, but much too thick carbon on SnO$_2$ nanoparticle may significantly reduce the catalytic activity. In our experiment, the oxygen-reduction-reaction activity of the gas sensing material increases firstly, and then decreases with the graphitized-carbon deposited on the SnO$_2$ nanoparticles increasing. It may be the combined effects of graphitization degree and thickness of carbon layer. The sample S3 (carbon-decorated SnO$_2$ nanoparticles prepared at 400 °C) shows the highest response to C$_2$H$_2$, that may illustrate the graphitization degree and thickness of carbon layer in sample S3 is optimal in this experiment.

Besides the crystal structure and thickness of carbon deposited on the $SnO_2$ nanoparticles, the carbon deposition process may induce the concentration of oxygen vacancies in $SnO_2$ nanoparticles to change, which may be affected by different sintering temperature. These factors also have a strong impact on the gas response performance of the sensors base on carbon-decorated $SnO_2$ nanoparticles. And the related work we will further study.

**4. Conclusion**

The high response and low-cost gas sensors were fabricated by the carbon-decorated $SnO_2$ nanoparticles, which was prepared by CVD method using $C_2H_2$ splitting decomposition. The gas-sensing performance of the sensors can be modulated by surface modification via changing preparation temperatures ($T_p$). Due to the different carbon-loading of $SnO_2$ nanoparticles, the response of sensors increases with the $T_p$ and reaches its maximum value of 106.3 at 400 $^{\circ}$C, then decreases when $T_p$ rose continually. The present result indicates the gas sensors based on carbon-decorated $SnO_2$ nanoparticles by CVD method has strong potential in preparation of gas sensing material with high response and low cost.

**Acknowledgement**

This work was supported by the Fundamental Research Funds for the Central Universities (ZYGX2014J087), Sichuan Province Science and Technology Support Project (2014GZ0151), and Sichuan Province Application Foundation Research Project (2015JY0066). We would like to express our appreciation for the helpful comments of Dr. Yunxiang Pan at Nanyang Technological University.